\documentstyle[12pt]{article}  \textwidth=15.5cm
 \renewcommand{\dag}{{\,\dagger}}
\newcommand{\ket}{\!\!>} \newcommand{\beq}{\begin{equation}}
\newcommand{\beqn}{\begin{eqnarray}} \newcommand{\eeq}{\end{equation}}
\newcommand{\eeqn}{\end{eqnarray}}

\begin{document} \bibliographystyle{plain}

\title{\vspace*{0.7cm} The structure of the spectrum of\\ anomalous
dimensions in the\\ $N$--vector model in $4-\epsilon$~dimensions
\vspace*{0.5cm}} \author{ Stefan~K.~Kehrein\thanks{E--mail:
ej6@vm.urz.uni-heidelberg.de}~~and Franz~Wegner \vspace*{0.2cm}\\
Institut f\"ur Theoretische Physik,\\
Ruprecht--Karls--Universit\"at,\\ D-69120~Heidelberg, F.R.~Germany
\vspace*{0.5cm} } \date{March 6, 1994} \maketitle \begin{abstract} In
a recent publication we have investigated the spectrum of anomalous
dimensions for arbitrary composite operators in the critical
$N$--vector model in $4-\epsilon$~dimensions.  We could establish
properties like upper and lower bounds for the anomalous dimensions in
one--loop order.  In this paper we extend these results and explicitly
derive parts of the one--loop spectrum of anomalous dimensions.  This
analysis becomes possible by an explicit representation of the
conformal symmetry group on the operator algebra.  Still the structure
of the spectrum of anomalous dimensions is quite complicated and does
generally not resemble the algebraic structures familiar from two
dimensional conformal field theories.  \vspace*{1cm} \end{abstract}

\section{Introduction} The renormalization of composite operators is a
classical problem in quantum field theory \cite{zimmermann73a}.  In a
recent publication \cite{kehrein93a} we have investigated some aspects
of this problem in the $N$--vector model in $4-\epsilon$~dimensions.
We found a one--loop order solution for the spectrum of anomalous
dimensions in terms of a ``two--particle" interaction operator~$V$
acting on a Hilbert space~$\cal C$ isomorphic to the space of
composite operators.  This hermitean operator~$V$ encodes all the
information about the one--loop spectrum of anomalous dimensions and
the corresponding scaling eigenoperators.  Therefore one is interested
to know as much as possible about the explicit structure of the
spectrum of~$V$.  However in ref.~\cite{kehrein93a} we have only been
able to diagonalize~$V$ explicitly for composite operators consisting
of two or three elementary fields.  In this paper we extend these
results and derive anomalous dimensions for larger classes of
composite operators.

A complete classification of the spectrum of anomalous dimensions is
possible in two dimensional conformal field theories.  It is an
interesting question to find out how much of the algebraic structures
there can also be found in $d$~dimensional conformal field theories
for $d>2$.  But it is well known that beyond two dimensions conformal
symmetry yields less stringent conditions.  Therefore many questions
in $d$~dimensional conformal field theory still remain open.
Non--perturbative treatments generally generate complicated series of
equations that cannot be easily solved.

In contrast we are working within the framework of one--loop order
perturbation theory.  It is in principle a straightforward problem to
diagonalize our operator~$V$ and to calculate anomalous dimensions for
whatever composite operators one is interested in.  Conformal symmetry
enters here only as the symmetry group of~$V$:  One can restrict
attention to conformal invariant operators (CIOs) as is well known
from conformal field theory.  However the dimensions of the vector
spaces of mixing CIOs can become arbitrarily large for a large number
of gradients in the composite operators.  Thus the well--known problem
of operator mixing occurs.  One has to resort to numerical
diagonalizations of~$V$ and finds a rather complicated spectrum of
non--rational anomalous dimensions.  This spectrum seems to show
little resemblance to the familiar algebraic structures from two
dimensional conformal field theories.

Nevertheless some properties of the spectrum can be understood
analytically.  It emerges a picture that in many spaces of mixing CIOs
there are operators with smallest and rational anomalous dimensions
(``ground states"), and ``excitations" with typically non--rational
eigenvalues above.  The anomalous dimensions of the ``ground states"
can be derived in closed form.  In particular for a sufficiently large
number of gradients there is a highly degenerate subspace of CIOs with
vanishing anomalous dimensions in one--loop order.  This agrees with
Parisi's intuitive conjecture that a high spin ``effectively"
separates space--time points \cite{callan73a}.

The structure of this paper is as follows.  Sects.~2 and~3 give a
short summary of the main results of our previous publication
\cite{kehrein93a} and define our notation.  In sect.~2 the
operator~$V$ is introduced that generates the spectrum of anomalous
dimensions.  We also present the explicit representation of the
conformal symmetry group on the operator algebra.  Sect.~3 is
concerned with the explicit diagonalization of~$V$ for composite
operators consisting of $n=2$ and $n=3$~fields.  Some remarks on the
$N\rightarrow\infty$ limit of the spectrum follow in sect.~4.
Explicit values for the anomalous dimensions can be given if the
composite operators contain less than six gradients.  This is shown in
sect.~5.  In sect.~6 some improved lower limits for the general
spectrum problem are established.

Sects.~7 and 8 are concerned with the more complicated parts of the
spectrum of anomalous dimensions for a large number of gradients.  In
sect.~7 we allow for arbitrary SO(4) spin structure, whereas the
important class of composite operators with $l$~gradients and SO(4)
spin structure $(l/2,l/2)$ is treated in more detail in sect.~8.  In
work on conformal field theory one frequently considers this spin
structure only since only this type of CIOs appears in the operator
product expansion of two scalar operators (see ref.~\cite{mack77a}).
One can view the spectrum problem for such CIOs as an energy spectrum
for a certain quantum mechanical system of bosons interacting with
$\delta$--potentials.  Finally sect.~9 deals with the particularly
intriguing highly degenerate class of CIOs with vanishing anomalous
dimenisions in one--loop order.  A short summary of the results is
presented in sect.~10.

\section{The spectrum--generating operator $V$} We briefly sum up the
main results of our earlier paper \cite{kehrein93a}.  An attempt is
made to make the presentation here self--consistent, for more details
the reader is referred to the earlier work.

The $N$--vector model of the field $\vec\phi=(\phi_1,\ldots,\phi_n)$
in $4-\epsilon$~dimensions is governed by the Lagrangian \beq {\cal
L}=\frac{1}{2}\:(\partial\vec\phi)^2+\frac{1}{2}\:m_0^2\:\vec\phi^2+
\frac{g_0}{4!}\:
(\vec\phi^2)^2 .  \eeq Composite local operators in this model
consisting of $n$~elementary fields and gradients acting on these
fields are represented as polynomials of
fields~$\Phi_i^{(j;m_1,m_2)}$, $j=0,1,\ldots$ and
$m_1,m_2=-j/2,j/2+1,\ldots,j/2$.  Here $\Phi_i^{(j;m_1,m_2)}$ is an
elementary field~$\phi_i$ with $j$~gradients \beq
\Phi_i^{(j;m_1,m_2)}\stackrel{\rm def}{=}
h_{\alpha_1\ldots\alpha_j}^{(m_1,m_2)}\:
\partial_{\alpha_1}\ldots\partial_{\alpha_j}\:\phi_i(0) , \eeq where
$h_{\alpha_1\ldots\alpha_j}^{(m_1,m_2)}$ is a symmetric and traceless
SO(4) tensor belonging to the irreducible representation $(j/2,j/2)$
of SO(4).  $h_{\alpha_1\ldots\alpha_j}^{(m_1,m_2)}$ can be constructed
via the one to one correspondence with homogeneous harmonic
polynomials of degree~$j$ \beq
H_j^{(m_1,m_2)}(x)=h_{\alpha_1\ldots\alpha_j}^{(m_1,m_2)}\:
x_{\alpha_1}\cdot\ldots\cdot x_{\alpha_j} \eeq with the generating
functional \beq (u\,\cdot\, x)^j=\sum_{m_1,m_2=-j/2}^{j/2} 2^j \,j!\:
\sqrt{{j \choose m_1+j/2}\:{j \choose m_2+j/2}}\:
H_j^{(m_1,m_2)}(x)\,\cdot\,t^{m_1+j/2}\,s^{m_2+j/2} \eeq and \beq
u=(i-i\,t\,s,-i\,t-i\,s,-t+s,1+t\,s) .  \eeq We define a creation
operator $a_i^{(j;m_1,m_2)\dag}$ corresponding to multiplication with
the field $\Phi_i^{(j;m_1,m_2)}$ and an annihilation operator
$a_i^{(j;m_1,m_2)}$ corresponding to the derivative
$\partial/\partial\Phi_i^{(j;m_1,m_2)}$ with Bose commutation
relations \beq [a_i^{(j;m_1,m_2)},a_{i'}^{(j';m_1',m_2')\dag}]
=\delta_{ii'}\,\delta_{jj'}\,\delta_{m_1m_1'}\,\delta_{m_2m_2'} .
\eeq $a_i^{(j;m_1,m_2)}$ and $a_{i}^{(j;m_1,m_2)\dag}$ operate on a
Hilbert space~$\cal C$ with vacuum~$|\Omega\ket$ that can therefore be
thought of as the space of composite operators.  Notice that by
omitting terms like $\Delta\phi_i$ in our construction of~$\cal C$ we
have removed a class of redundant operators that is of no physical
interest (compare ref.~\cite{kehrein93a}).

As we have shown in our first paper, scaling eigenoperators and
anomalous dimensions of the critical model are in one--loop order
eigenvectors and eigenvalues of a two--particle interaction operator
\beqn V_N&=&\frac{1}{2}\:\sum_{d,d',e,e'=1}^N\sum_Q\:
v_Q\;\cdot\;(\delta_{ee'}\,\delta_{dd'}+\delta_{ed}\,\delta_{e'd'}
+\delta_{ed'}\,\delta_{e'd}) \label{eq22} \\ &&\qquad\qquad\times\quad
a_e^{(j;m_1,m_2)\dag}\,a_{e'}^{(j';m_1',m_2')\dag}\,
a_d^{(l;n_1,n_2)}\,a_{d'}^{(l';n_1',n_2')} .  \nonumber \eeqn The sum
$\sum_Q$ runs over all SO(4) quantum numbers of the creation and
annihilation operators.  The interaction kernel consists of a product
of four SO(3) Clebsch--Gordan coefficients \beqn
v_Q&=&\delta_{l+l',j+j'}\:\delta_{n_1+n_1',m_1+m_1'}\:
\delta_{n_2+n_2',m_2+m_2'}
\label{eq23} \\
&&\times\sum\limits_{k=0}^{\left[\frac{l+l'}{2}\right]}
\frac{1}{l+l'-2k+1}\; (l/2,n_1;l'/2,n_1'\:|\:(l+l')/2-k,n_1+n_1')
\nonumber \\ &&\hspace*{3.8cm}\times\,
(l/2,n_2;l'/2,n_2'\:|\:(l+l')/2-k,n_2+n_2') \nonumber\\
&&\hspace*{3.8cm}\times\,
(j/2,m_1;j'/2,m_1'\:|\:(j+j')/2-k,m_1+m_1')\nonumber\\
&&\hspace*{3.8cm}\times\,
(j/2,m_2;j'/2,m_2'\:|\:(j+j')/2-k,m_2+m_2')\nonumber \eeqn Obviously
operator mixing is here in one--loop order restricted to composite
operators with the same number of fields and gradients.

An eigenvector $|\psi\ket$ of $V_N$ with $n$~fields and $l$~gradients
\beq V_N\:|\psi\ket = \alpha\:|\psi\ket \eeq corresponds to an
eigenoperator in one--loop order with anomalous dimension \beq
\lambda=\epsilon\:\cdot\:\frac{\alpha}{N+8} + {\rm O}(\epsilon^2).
\eeq The critical exponent of this eigenoperator is \beq x=
l+n\:\left(1-\frac{\epsilon}{2} \right) + \lambda \eeq and the full
scaling dimension \beq y=d-x=
4-\epsilon-l-n\:\left(1-\frac{\epsilon}{2}\right) - \lambda .
\label{eq27} \eeq Operators are relevant, marginal or irrelevant for
$y>0$, $y=0$, $y<0$ respectively.

In the sequel we will frequently restrict ourselves to the scalar
$\phi^4$--theory \linebreak[3] ($N=1$) or to operators that are
completely symmetric and traceless with respect to the O($N$) indices.
In both cases one can omit the $N$--vector indices of the creation and
annihilation operators and simply write instead of (\ref{eq22}) \beq
V=\frac{1}{2}\:\sum_Q\:v_Q\;
a^{(j;m_1,m_2)\dag}\,a^{(j';m_1',m_2')\dag}\,a^{(l;n_1,n_2)}\,
a^{(l';n_1',n_2')} .  \label{eq24} \eeq
Anomalous dimensions for eigenoperators with
eigenvalue~$\alpha$ are then \beq \lambda=
\epsilon\,\cdot\,\mu\,\cdot\,\alpha+{\rm O}(\epsilon^2) .
\label{eq25} \eeq with \beq \mu=\left\{\begin{array}{l}1/3 \quad {\rm
for~} N=1 \\ 2/(N+8) \quad \mbox{for O($N$) symmetric and traceless
tensors}.  \end{array} \right.  \label{eq26} \eeq Eigenoperators in
the scalar theory therefore generate O($N$) symmetric and traceless
eigenoperators \beqn \lefteqn{
\Phi^{(j^{(1)};m^{(1)}_1,m^{(1)}_2)}\,\cdot\,\ldots\,\cdot\,
\Phi^{(j^{(n)};m^{(n)}_1,m^{(n)}_2)} + \ldots} \label{eq26a} \\
&\longrightarrow& t_{i_1\ldots i_n}\:\left(
\Phi_{i_1}^{(j^{(1)};m^{(1)}_1,m^{(1)}_2)}\,\cdot\,\ldots\,\cdot\,
\Phi_{i_n}^{(j^{(n)};m^{(n)}_1,m^{(n)}_2)} + \ldots\right) , \nonumber
\eeqn where $t_{i_1\ldots i_n}$ is a symmetric and traceless tensor.
Notice that all O($N$) totally antisymmetric composite operators are
annihilated by~$V_N$ and have vanishing anomalous dimensions in
one--loop order.  We will ignore this uninteresting class of operators
in the sequel.

It has already been mentioned in our previous paper that $V_N$ is a
hermitean positive semi--definite operator.  Hence its eigenvalues are
real positive numbers that can only make canonically irrelevant
operators even more irrelevant according to (\ref{eq27}).  In contrast
to some $2+\epsilon$~expansions
\cite{kravtsov88a,kravtsov89a,wegner90a,wegner91a,mall92a,castilla93a}
the stability of the nontrivial fixed point in $4-\epsilon$~dimensions
is in one--loop order not endangered by high--gradient operators.

The spatial symmetry group of $V_N$ is a representation of the
conformal group in four dimensions~O(5,1) on the operator
algebra~$\cal C$.  Conformal symmetry should of course be expected at
the critical point of a second order phase transition.  The
15~generators of SO(5,1) can conveniently be expressed as \beqn
\partial_{st}&=&\sum_{i=1}^N \sum_{(j;m_1,m_2)}
\sqrt{(j/2+1+s\,m_1)\,(j/2+1+t\,m_2)}\:a_i^{(j+1;m_1+s/2,m_2+t/2)\dag}\:
a_i^{(j;m_1,m_2)} \nonumber \\ \partial_{st}^\dag&=&\sum_{i=1}^N
\sum_{(j;m_1,m_2)}
\sqrt{(j/2+1+s\,m_1)\,(j/2+1+t\,m_2)}\:a_i^{(j;m_1,m_2)\dag}\:
a_i^{(j+1;m_1+s/2,m_2+t/2)} \nonumber \\ J_s^1&=&\sum_{i=1}^N
\sum_{(j;m_1,m_2)}
\sqrt{(j/2-s\,m_1)\,(j/2+1+s\,m_1)}\:a_i^{(j;m_1+s,m_2)\dag}\:
a_i^{(j;m_1,m_2)} \nonumber \\ J_z^1&=&\sum_{i=1}^N \sum_{(j;m_1,m_2)}
m_1\;a_i^{(j;m_1,m_2)\dag}\:a_i^{(j;m_1,m_2)} \label{eq28} \\
J_t^2&=&\sum_{i=1}^N \sum_{(j;m_1,m_2)}
\sqrt{(j/2-t\,m_2)\,(j/2+1+t\,m_2)}\:a_i^{(j;m_1,m_2+t)\dag}\:
a_i^{(j;m_1,m_2)} \nonumber \\ J_z^2&=&\sum_{i=1}^N \sum_{(j;m_1,m_2)}
m_2\;a_i^{(j;m_1,m_2)\dag}\:a_i^{(j;m_1,m_2)} \nonumber \\
J_0&=&\sum_{i=1}^N \sum_{(j;m_1,m_2)}
(j/2+1/2)\;\:a_i^{(j;m_1,m_2)\dag}\:a_i^{(j;m_1,m_2)} \nonumber \eeqn
with $s,t=-1,+1$.  For later purposes we also introduce the linear
combinations \beq \begin{array}{llll}
J_{++}^1=J_0+J_z^1,&J_{--}^1=J_0-J_z^1,&J_{+-}^1=J_+^1,&J_{-+}^1=J_-^1,\\
J_{++}^2=J_0+J_z^2,&J_{--}^2=J_0-J_z^2,&J_{+-}^2=J_+^2,&J_{-+}^2=J_-^2.
\end{array} \label{eq29} \eeq Now it is clearly sufficient to restrict
our attention to conformal invariant operators (CIOs) annihilated by
the generators of special conformal transformations \beq
\partial_{st}^\dag\:|\psi\ket = 0 \qquad \forall s,t=-1,+1 \eeq since
all derivative operators $\partial_{s_1 t_1}\ldots\partial_{s_p
t_p}\,|\psi\ket$ reproduce the anomalous dimension of~$|\psi\ket$.
Considering only CIOs reduces the dimensions of the spaces of mixing
operators considerably and renders the diagonalization problem of
$V_N$ more tractable.

The vector space of CIOs with $n$~fields, $l$~gradients and spin
structure~$(j_1,j_2)$ in the two SO(3) sectors of SO(4) will be
denoted by~${\cal C}[n,(l;j_1,j_2)]$, where $j_1,j_2=l/2, l/2-1,
\ldots, 1/2{\rm~or~}0$.  Since eq.~(\ref{eq28}) gives an explicit
representation of the generators of special conformal transformations
$\partial_{st}^\dag$, it is in principle straightforward to construct
CIOs in a given ${\cal C}[n,(l;j_1,j_2)]$.  If one applies a
generator~$\partial_{st}^\dag$ on a not conformally invariant
operator, the result is a composite operator with one gradient less
than the original operator.  Thus repeated application of the
generators of special conformal transformations finally yields a
conformal invariant operator that is mapped to zero by
all~$\partial_{st}^\dag$.  This immediately provides a method for
constructing CIOs that is simple to implement if the dimension of a
space of CIOs is small.

\begin{description} \item[\underline{Example:}] \ \newline For $n=2$
fields one finds e.g.  exactly one O($N$) scalar CIO $T^{(l)}$ with an
even number of fields~$l$.  $T^{(l)}$ transforms as a tensor with spin
structure $(l/2,l/2)$ under SO(4)--rotations.  Its components
$T^{(l)(m_1,m_2)}$ \beq J_z^{1,2}\:  T^{(l)(m_1,m_2)} =
m_{1,2}\:T^{(l)(m_1,m_2)} \eeq can be generated from
$T^{(l)(l/2,l/2)}$ by application of $J_-^{1,2}$ \beq
T^{(l)(m_1,m_2)}=\frac{(l/2+m_1)!\,(l/2+m_2)!}{l!^2\,(l-2m_1-1)!!\,
(l-2m_2-1)!!}\:
\left(J_-^1\right)^{l/2-m_1}\,\left(J_-^2\right)^{l/2-m_2}\:
T^{(l)(l/2,l/2)}
\eeq where \beq T^{(l)(l/2,l/2)}=\frac{1}{2}\sum_{k=0}^l (-1)^k\,{l
\choose k}\:
\vec\Phi^{(k;k/2,k/2)}\cdot\vec\Phi^{(l-k;\frac{l-k}{2},\frac{l-k}{2})} .
\eeq Notice that $T^{(2)}$ is proportional to the stress tensor.
\end{description} Usually we do not explicitly write down the form of
the scaling eigenoperators in the sequel.

If dim~${\cal C}[n,(l;j_1,j_2)]$ is large the problem of finding all
CIOs is harder.  We come back to this problem in sect.~7.  For the
important case ${\cal C}[n,(l;l/2,l/2)]$ we give a complete and
constructive solution in sect.~8.

\section{Conformal invariant operators with $n\leq 3$ fields} For two
and three fields the spectrum of CIOs has already been worked out in
our previous paper~\cite{kehrein93a}.  The results are summed up in
tables~1 to 3.  All the CIOs in these tables have SO(4) spin structure
$(l/2,l/2)$.  \newsavebox{\k}
\sbox{\k}{\framebox[5mm]{\rule{0mm}{2.7mm}}}
\renewcommand{\arraystretch}{1.2}

However for $n=3$ fields one can in general also construct CIOs with a
different SO(4) spin structure, or more than one CIO with spin
structure $(l/2,l/2)$.  Since the spectrum--generating operator~$V$
acts like a projection operator (except for a slight complication for
the last case in table~3) on the three particle space, these
additional CIOs always have vanishing anomalous dimensions in
order~$\epsilon$.  In general it is quite difficult to give an
explicit formula for the number of these additional CIOs.  At least
for $N=1$ (or O($N$) completely symmetric and traceless tensors) with
SO(4) spin structure $(l/2,l/2)$, we will establish the following
generating function in sect.~7 \beq \sum_{l=0}^\infty \left({\rm
dim}\:{\cal C}[n=3,(l;l/2,l/2)] \right)\,\cdot\, x^l =
\frac{1}{(1-x^2)(1-x^3)} .  \eeq Thus there are $\left({\rm
dim}\:{\cal C}[n=3,(l;l/2,l/2)]-1\right)$ CIOs with zero eigenvalues
besides the CIOs with non--zero eigenvalues from tables~2 and~3.  For
large~$l$ this number increases approximately linearly with~$l$,
therefore vanishing anomalous dimensions dominate the spectrum for
$l\rightarrow\infty$.

Let us mention that we have compared our anomalous dimensions for CIOs
with Young frame~~\usebox{\k}\usebox{\k}\usebox{\k}~~for
$l=0,2,3,4,5,6$ against results obtained by Lang and R\"uhl in a
$1/N$--expansion \cite{ruehl94a}.  Their results give the exact
dependence on the dimension $2<d<4$ in order $1/N$ and can therefore
be expanded in~$\epsilon$ for $d=4-\epsilon$.  Whereas our results are
obtained in first order in~$\epsilon$ but give the full
$N$--dependence.  We have found complete agreement in
order~$\epsilon/N$.

\section{The large-$N$ limit} One can easily solve the large $N$ limit
of the spectrum problem for $V_N$.  This serves only as another
consistency check of our approach since the order~O($N^0$)
contributions of the anomalous dimensions are trivial in a
$1/N$--expansion.

First one notices that the action of~$V_N$ on a composite
operator~$|\psi_i\ket$ can be split up in a term linear in~$N$ and a
constant term \beq V_N\,|\psi_i\ket=N\,\cdot\,\sum_j
\gamma_{ij}\,|\psi_j\ket + \sum_{j}\gamma_{ij}'\,|\psi_j\ket ,
\label{largeN0} \eeq where the mixing matrices $\gamma_{ij},
\gamma_{ij}'$ are independent of~$N$ and all composite operators
normalized.  The O($N^0$) term~$\epsilon\cdot\alpha$ in the anomalous
dimension of an eigenoperator in this double limit ($d\rightarrow 4,
N\rightarrow\infty$) \beq \lambda = \epsilon\,\alpha + {\rm O}(1/N) +
{\rm O}(\epsilon^2) .  \label{largeN2} \eeq is simply the respective
eigenvalue of $\gamma_{ij}$.  But the terms linear in $N$ in
eq.~(\ref{largeN0}) come solely from~$V_N$ acting on composite
operators like \beq S=\prod_{i=1}^k \left(\sum_Q c_Q\:
\vec\Phi^{(j^{(i)};m_1^{(i)},m_2^{(i)})}\,\cdot\,
\vec\Phi^{(j^{(i)'};m_1^{(i)'},m_2^{(i)'})} \right) \label{largeN1}
\eeq with
$Q=(j^{(i)};m_1^{(i)},m_2^{(i)}),(j^{(i)'};m_1^{(i)'},m_2^{(i)'})$ and
\beq V_N\,(S) = \prod_{i=1}^k V_N\left(\sum_Q c_Q\:
\vec\Phi^{(j^{(i)};m_1^{(i)},m_2^{(i)})}\,\cdot\,
\vec\Phi^{(j^{(i)'};m_1^{(i)'},m_2^{(i)'})} \right) + {\rm O}(N^0) .
\label{eqN2} \eeq Now $V_N$ acts as a projection operator on each
product in (\ref{eqN2}) \beq V_N\left(
\vec\Phi^{(j^{(i)};m_1^{(i)},m_2^{(i)})}\,\cdot\,
\vec\Phi^{(j^{(i)'};m_1^{(i)'},m_2^{(i)'})} \right) \propto
p(\vec\partial)\:(\vec\phi^2) \eeq where $p(\vec\partial)$ is a
certain homogeneous polynomial of degree~$j^{(i)}+j^{(i)'}$ in the
derviatives.  Trivially we have for all such homogeneous polynomials
\beq
V_N\left(p(\vec\partial)\,\vec\phi^2\right)=(N+2)\:p(\vec\partial)\,
\vec\phi^2 .  \eeq Therefore eigenoperators in the double limit
$d\rightarrow 4,
N\rightarrow\infty$ are of type \beq
O=\left[p_1(\vec\partial)\,\vec\phi^2\right]\,\left[p_2(\vec\partial)\,
\vec\phi^2\right]\,
\cdot\ldots\cdot\,\left[p_{n_0}(\vec\partial)\,\vec\phi^2\right]\:\cdot\:S ,
\eeq where each term of $S$ from eq.~(\ref{largeN1}) \beq
\left(\sum_Q c_Q\:  \vec\Phi^{(j^{(i)};m_1^{(i)},m_2^{(i)})}\,\cdot\,
\vec\Phi^{(j^{(i)'};m_1^{(i)'},m_2^{(i)'})} \right) \eeq is as a
vector in $\cal C$ orthogonal to $\vec\phi^2$ and all total
derivatives thereof.  Also each product of~$O$ with a traceless
composite operator $C_{i_1\ldots i_p}$ consisting of $p$~elementary
fields~$\Phi_i^{(j;m_1,m_2)}$ is an eigenoperator in this double
limit.  The anomalous dimension of~$O$ depends only on the
number~$n_0$ of factors~$\vec\phi^2$ or total derivatives of
$\vec\phi^2$ \beqn \lambda&=&\epsilon\,\cdot\,n_0+{\rm O}(1/N) +{\rm
O}(\epsilon^2) \\ \Longrightarrow {\rm~Critical~exponent~{}}
x&\!\!\!=&(n-2n_0)\cdot \left(\frac{d}{2}-1\right)+l+2\,n_0+{\rm
O}(1/N) +{\rm O}(\epsilon^2) .  \nonumber \eeqn As should be expected
this is consistent with the spherical model limit.

It is, however, not too surprising that the $1/N$--corrections in
(\ref{largeN2}) are very involved.  In fact a $1/N$--expansion beyond
the zeroth order term does here not provide any additional
simplifications and is already equivalent to the full problem of
diagonalizing $V_N$.

In the remainder of this paper we therefore keep the exact
$N$--dependence.  But since we are mainly interested in the spatial
symmetry, we restrict the discussion to O($N$) tensors consisting of
$n$~elementary fields with the Young frame \begin{equation}
\framebox[5mm]{1}\framebox[20mm]{\ldots\rule{0mm}{2.7mm}}
\framebox[5mm]{$n$\rule{0mm}{2.7mm}}~~, \label{largeN3} \end{equation}
that is to O($N$) completely symmetric and traceless tensors.  By
virtue of eq.~(\ref{eq26}) we need thus in fact only discuss the
spectrum of anomalous dimensions in the scalar $\phi^4$--theory:
Eq.~(\ref{eq26a}) yields the generalization from $N=1$ to O($N$)
tensors of type (\ref{largeN3}) and the anomalous dimensions are
related according to eq.~(\ref{eq26}).

\section{Conformal invariant operators with $l\leq 5$ gradients} As
already mentioned, operator mixing due to the increasing number of
conformal invariant operators makes the diagonalization problem in
general intractable for a large number of gradients~$l$.  There are
typically no ``simple" rational anomalous dimensions besides the ones
belonging to the lower bound as established in the next section.

For the scalar $\phi^4$--theory or O($N$) tensor with the
transformation properties (\ref{largeN3}), however, the spaces of
mixing CIOs ${\cal C}[n,(l;j_1,j_2)]$ have maximum dimension two for
$l\leq 5$~gradients.  All eigenvalues can easily be derived explicitly
in this case.  For the sake of completeness these eigenvalues~$\alpha$
are summed up in table~4.  The corresponding anomalous dimensions
follow from eq.~(\ref{eq26}).  Again we have checked a few randomly
chosen anomalous dimensions against results obtained in the
$1/N$--expansion by Lang and R\"uhl \cite{ruehl93a,ruehl94a} and
always found consistency.

\section{Lower limits for the spectrum} Since the spectrum--generating
operator~$V$ is positive semi--definite as shown in our previous
paper~\cite{kehrein93a}, zero is an obvious lower limit for the
eigenvalues~$\alpha$ of~$V$ and hence for the anomalous
dimensions~$\lambda$ in (\ref{eq27}).  However zero is not a strict
lower limit for the spectrum and can be improved for n larger than
approximately $l/2$.

In order to establish a more stringent lower bound one first notices
that \beq C=\sum_{s,t} \partial_{st}\,\partial_{st}^\dag - \frac{1}{2}
\sum_{s,t} J_{st}^{1} \,J_{ts}^{1} -\frac{1}{2} \sum_{s,t}
J_{st}^{2}\,J_{ts}^{2} +4J_0 \eeq is a Casimir operator of the group
SO(5,1) from (\ref{eq28}).  Then \beq L=\frac{1}{6}\:\left(
C+\frac{7}{2} \hat{n}(\hat{n}-1)-\frac{3}{2} \hat{n} \right) \eeq also
commutes with all generators of SO(5,1) and with the two--particle
interaction operator~$V$ from (\ref{eq24}).  Here $\hat{n}$ is an
operator that counts the number of fields \beq \hat{n}=
\sum_{(j;m_1,m_2)} \:  a^{(j;m_1,m_2)\dag}\,a^{(j;m_1,m_2)} .  \eeq
$V$ and $L$ can be simultaneously diagonalized and since
$[L,\partial_{st}]=0$ we need only consider CIOs as usual.  Acting on
${\cal C}[n,(l;j_1,j_2)]$ the Casimir operator~$C$ is equal to \beq
-j_1\,(j_1+1)-j_2\,(j_2+1)-\frac{(n+l)^2}{2}+2\,(n+l) , \eeq therefore
\beq L\Big|_{{\cal C}[n,(l;j_1,j_2)]} = \frac{1}{2}
n\,(n-1)-\frac{1}{6} n\,l -\frac{1}{12} l\,(l-4) -\frac{1}{6}
j_1\,(j_1+1) -\frac{1}{6} j_2\,(j_2+1) .  \label{eq51} \eeq Eq.
(\ref{eq51}) represents the lower limit that we want to establish:
Because of \beq L\:\Phi^{(0;0,0)} = 0 \eeq $L$ is a genuine
two--particle operator.  But acting on states with $n=2$ (notice that
$n=2$ implies $l$ even and $j_1=j_2=l/2$) one finds $L=1, 0, -7/3,
\ldots$ for $l=0,2,4,\ldots$.  On the other hand $V=1$ for $n=2, l=0$
and $V=0$ otherwise.  Combining the properties above one immediately
sees \beq V - L \geq 0 \eeq or \beq V\Big|_{{\cal C}[n,(l;j_1,j_2)]}
\geq \max\left(0, \frac{1}{2} n\,(n-1)-\frac{1}{6} n\,l -\frac{1}{12}
l\,(l-4) -\frac{1}{6} j_1\,(j_1+1) -\frac{1}{6} j_2\,(j_2+1)\right) .
\label{lowbound} \eeq We can go a step further and show that for
$n\geq l\geq 4$ and spin structure $(j,j)$, that is equal spin in both
SO(3) sectors, there always exists an eigenoperator with an eigenvalue
given by eq.~(\ref{eq51}).  $L$ is thus a strict lower limit for this
class of composite operators.

Let us briefly sketch this proof of $L$ being a strict lower bound.
Essentially one has to show that for $n\geq l$ an operator exists in
${\cal C}[n,(l;j,j)]$ with eigenvalue~(\ref{eq51}).  Now in fact this
eigenoperator turns out to have the following general structure \beqn
O&=&(\Phi^{(0;0,0)})^{n-l}\;\sum_{\{\alpha_i\} }
O^{(0;0,0)}_{\alpha_1\ldots\alpha_4}\:
[\alpha_1,\alpha_2,\alpha_3,\alpha_4] \label{eq52} \\
&&+(\Phi^{(0;0,0)})^{n+1-l}\;\sum_{m_1,m_2=-1}^1 \sum_{\{\beta_i\} }
O^{(2;m_1,m_2)}_{\beta_1\ldots\beta_4}\:\Phi^{(2;m_1,m_2)}\:
[\beta_1,\beta_2,\beta_3,\beta_4] \nonumber \\ &&+ \mbox{~other terms
with higher powers of~}\Phi^{(0;0,0)} , \nonumber \eeqn where we have
used the compact notation \beq [\alpha_1,\alpha_2,\alpha_3,\alpha_4]
\stackrel{\rm def}{=} (\Phi^{(1;-1/2,-1/2)})^{\alpha_1}
(\Phi^{(1;-1/2,1/2)})^{\alpha_2} (\Phi^{(1;1/2,-1/2)})^{\alpha_3}
(\Phi^{(1;1/2,1/2)})^{\alpha_4} .  \eeq The important thing to notice
is that all the terms not written out in eq.~(\ref{eq52}) cannot be
mapped on $(\Phi^{(0;0,0)})^{n-l}\;\sum_{\{\alpha_i\} }\ldots$ by~$V$.
Hence we only need to know something about the coefficients
$O^{(0;0,0)}_{\alpha_1\ldots\alpha_4}$ and
$O^{(2;m_1,m_2)}_{\beta_1\ldots\beta_4}$.  This information can be
obtained by using $O\in {\cal C}[n,(l;j,j)]$, for example by requiring
\beq J^1_z| O\ket =J^2_z| O\ket = j\,|O\ket \label{eq53} \eeq \beq
\partial_{st}^\dag |O\ket =J^1_{+1}| O\ket =J^2_{+1}| O\ket = 0 .
\label{eq54} \eeq From eqs.~(\ref{eq53}) and (\ref{eq54}) one
concludes after some calculation \beq
O_{\alpha_1\ldots\alpha_4}^{(0;0,0)}=
\left\{\begin{array}{l}\displaystyle{ (-1)^\alpha\:{\displaystyle
l/2-j \choose \displaystyle \alpha} } \begin{array}{l}
\mbox{if~}\alpha=0,\ldots,l/2-j\mbox{~exists~with~}\\
\alpha_1=l/2-j-\alpha,\alpha_2=\alpha_3=\alpha,\alpha_4=l/2+j-\alpha
\end{array} \\ 0\mbox{~otherwise} \end{array} \right.  \eeq where we
have normalized $O^{(0;0,0)}_{l/2-j,0,0,l/2+j}=1$ and \beqn
2\,O^{(2;1,1)}_{l/2-j,0,0,l/2+j-2}+O^{(2;0,0)}_{l/2-j-1,0,0,l/2+j-1}
&=& -\frac{l}{2} -j \\
2\,O^{(2;-1,-1)}_{l/2-j-2,0,0,l/2+j}+O^{(2;0,0)}_{l/2-j-1,0,0,l/2+j-1}
&=& -\frac{l}{2} +j .  \nonumber \eeqn This information is sufficient
to derive \beqn V| O\ket&=&\left( \frac{1}{2} n\,(n-1)-\frac{1}{6}
n\,l -\frac{1}{12} l\,(l-4) -\frac{1}{3} j\,(j+1) \right) \nonumber\\
&&\qquad\times \left(\Phi^{(0;0,0)}\right)^{n-l}\:
\left[\frac{l}{2}-j,0,0,\frac{l}{2}+j\right] \\ &&+\mbox{~other
terms~}.  \nonumber \eeqn So if a CIO with structure (\ref{eq52})
exists, its eigenvalue must be \beq \alpha= \frac{1}{2}
n\,(n-1)-\frac{1}{6} n\,l -\frac{1}{12} l\,(l-4) -\frac{1}{3} j\,(j+1)
\eeq equal to the lower limit from eq.~(\ref{eq51}).

Finally one can show that CIOs of type (\ref{eq52}) with spin
structure $(l/2,l/2)$ exist for $n\geq l$, $l \neq 1$.  In the case of
non--maximum spin in the SO(3) sectors $(j,j), j\neq l/2$ they exist
for $n\geq l\geq 4$.

Thus the lower bound (\ref{eq51}) is strict in the classes:
\begin{itemize} \item ${\cal C}[n,(l;l/2,l/2)]$ for $n\geq l$, $l\neq
1$ \item ${\cal C}[n,(l;j,j)]$ for $n\geq l\geq 4$, $j\neq l/2$
\end{itemize}

Notice that the above proof cannot be extended to CIOs with unequal
spin $j_1\neq j_2$ in the SO(3) sectors.  This agrees with the fact
that in general (\ref{eq51}) is not a strict limit in this case as one
can see in the next section from explicit diagonalizations
of~$V$.\footnote{Essentially the proof fails because one can show
$O^{(0;0,0)}_{\alpha_1\ldots\alpha_4}\equiv 0$ for $j_1\neq j_2$.}

\section{More complex CIOs} Besides the lowest eigenvalues
established in the previous section, the spectrum of anomalous
dimensions consists mainly of irrational numbers for $n\geq4$,
$l\geq6$.  In order to investigate this spectrum, one first has to
construct the various CIOs in a given space ${\cal C}[n,(l;j_1,j_2)]$.
In the language of conformal field theory this amounts to determining
the field content of the theory.

Now the problem of constructing all CIOs is nontrivial for a large
number of gradients.  We have applied the following technique.
Operators $P_{st}$, $s,t=\pm 1$ are introduced \beq
P_{st}=\sum_{k=0}^\infty
\frac{1}{k!}\:\left(-\:\frac{a^{(1;s/2,t/2)\dag}}{a^{(0;0,0)\dag}}
\right)^k\:(\partial_{st}^\dag)^k \label{eqw2} \eeq with the following
properties \beqn [P_{st},P_{s't'}]&=&0 \nonumber \\ P_{st}^2&=&P_{st}
\label{eqw1} \\ \partial_{st}^\dag\,P_{s't'}&=&\left\{
\begin{array}{cl} P_{s't'}\,\partial_{st}^\dag &\quad\mbox{for~}s\neq
s'\mbox{~or~}t\neq t' \\ 0&\quad\mbox{for~}s=s'\mbox{~and~}t=t'
\end{array} \right.  \nonumber \eeqn Let us comment about the creation
operators $a^{(0;0,0)\dag}$ in the denominator of eq.~(\ref{eqw2})
later.  The construction of CIOs then runs as follows.  First a set of
states of~$|\psi'\ket$ is constructed by application of polynomials of
creation operators onto the vacuum without use of any operators
$a^{(1;s/2,t/2)\dag}$.  This is done is such a way that states with
the desired quantum numbers $[n,(l;j_1,j_2)]$ are obtained.  Then the
states \beq |\psi\ket=\left(\prod_{s,t=\pm 1}
P_{st}\right)\:|\psi'\ket \label{eqw3} \eeq are conformal invariant
due to the property (\ref{eqw1}).  Note that for a given~$l$ it is
sufficient to sum up to $k=l$ in eq.~(\ref{eqw2}).  If $n\geq l$ then
obviously no negative powers of $a^{(0;0,0)\dag}$ appear
in~$|\psi\ket$.  However since polynomials of~$a^{(1;s/2,t/2)\dag}$
generate only states with $j_1=j_2$, it is in fact sufficient that
$n\geq[l-1/2\,|j_1-j_2|]$.  Here $[l-1/2\,|j_1-j_2|]$ is the integer
part of $l-1/2\,|j_1-j_2|$.  Since $a^{(0;0,0)\dag}$ commutes with
$\partial_{st}^\dag$ one may multiply the conformal invariant states
with arbitrary powers of $a^{(0;0,0)\dag}$.

If the original states $|\psi'\ket$ are linearly independent then also
the states~$|\psi\ket$ have this property.  This is due to the fact
that $|\psi\ket$ and $|\psi'\ket$ differ only by states which contain
at least one factor~$a^{(1;s/2,t/2)\dag}$.  On the other hand a sum of
products each of which contains at least one factor
$a^{(1;s/2,t/2)\dag}$ acting on the vacuum is not conformally
invariant.  Suppose the term with the smallest number of factors
$a^{(1;s/2,t/2)\dag}$ has $k$ such factors.  Then at least one of the
operators obtained by application of $\partial_{s't'}^\dag$ contains a
term with $k-1$ such factors.  Thus the space ${\cal
C}[n,(l;j_1,j_2)]$ for $n\geq[l-1/2\,|j_1-j_2|]$ is spanned by the
operators~$|\psi\ket$ from eq.~(\ref{eqw3}).

We now come to the numerical solution of the eigenvalue problem
of~$V$.  One finds that generally the only ``simple" eigenvalues can
be written like $\alpha=1/2\,n(n-1)+n\,a+b$ with rational numbers $a$
and $b$.  Besides the trivial quadratic dependence on the number of
fields, these eigenvalues are linear functions of~$n$.  Notice that
the ground state eigenvalues (\ref{eq51}) are of just this linear
type.  We have therefore solved the eigenvalue problem for a general
number of fields $n\geq[l-1/2\,|j_1-j_2|]$ for $l\leq 8$, but only
present the eigenvalues of this simple type in tables~5 and~6.  That
is we list the degeneracy of the ground state eigenvalues and
additional linear eigenvalues.

The actual calculation was done as follows.  It is not necessary to
construct the conformal invariant operators~$|\psi\ket$ from
eq.~(\ref{eqw3}).  Instead one leaves out all contributions which
contain $a^{(1;s/2,t/2)\dag}$ in~$V$ and applies this to $|\psi'\ket$.
Similarly the part of~$V$ which destroys $\Phi^{(1;s/2,t/2)}$ is
applied to \beq
-\,\frac{a^{(1;s/2,t/2)\dag}}{a^{(0;0,0)\dag}}\,\partial_{st}^\dag\,|
\psi'\ket
\eeq and
$a^{(2;(s+s')/2,(t+t')/2)\dag}a^{(0;0,0)\dag}a^{(1;s/2,t/2)}
a^{(1;s'/2,t'/2)}$
to \beq
\frac{1}{2}\,\frac{a^{(1;s/2,t/2)^\dag}a^{(1;s'/2,t'/2)\dag}}{
a^{(0;0,0)\dag}a^{(0;0,0)\dag}}\,
\partial_{st}^\dag\partial_{s't'}^\dag\,|\psi'\ket .  \eeq This saves
considerable memory and computing time and still allows us to
determine the eigenvalues.  In this way we loose explicit hermitecity,
but since the calculation was done for general~$n$, it would be
difficult to make use of this property anyway.  Finally the
eigenvalues of $1/2\,n(n-1)+n\,A+B$ with matrices $A$ and $B$ have
been determined.

Table 5 lists the field content of our model for $l\leq 8$~gradients
and for\linebreak $n\geq[l-1/2\,|j_1-j_2|]$~fields constructed via
eq.~(\ref{eqw3}).  The additional quantum number~$\pi$ for $j_1=j_2$
corresponds to the discrete symmetry that interchanges the two SO(3)
representations \beq a^{(j;m_1,m_2)} \quad\longleftrightarrow\quad
a^{(j;m_2,m_1)} .  \eeq CIOs can be symmetric ($\pi=+$) or
antisymmetric ($\pi=-$) with respect to this symmetry.  Obviously
$\pi$ makes only sense for operators with the same spin in both SO(3)
sectors $j_1=j_2$.  From table~5 one sees that some ground state
eigenvalues occur with remarkably high degeneracies.  For example in
${\cal C}[n,(8;1,1,+)]$ one finds a fourfold degeneracy (in ${\cal
C}[n,(10;1,1,+)]$ the degeneracy is even fivefold).  We know no
explanation for this feature.  In table~6 the additional eigenvalues
of linear type are listed for $l\leq 8$.  If one goes to an even
larger number of gradients, the number of these additional linear
eigenvalues decreases again.  There might only be a finite set of
them though we cannot prove that.

\section{Spatially symmetric and traceless tensors} The simplest type
of conformal invariant operators that are spatially symmetric and
traceless tensors is studied in more detail in this section.  This
class of CIOs ${\cal C}[n,(l;l/2,l/2)]$ is especially important since
it is the only type that appears in an operator product expansion of
two scalar operators \cite{mack77a}.  In particular we can go beyond
the results of the last section and derive all CIOs in closed form.

First one notices that a composite operator in ${\cal
C}[n,(l;l/2,l/2)]$ can be expressed as \beq h_{\alpha_1\ldots
\alpha_l}^{(m_1,m_2)}\;
\left(\partial_{\alpha_1}\ldots\partial_{\alpha_{j_1}}\phi\cdot\ldots\cdot
\partial_{\alpha_{l-j_1+1}}\ldots\partial_{\alpha_l}\phi + \ldots
\right) \label{eq41} \eeq with a symmetric and traceless tensor
$h_{\alpha_1\ldots\alpha_l}^{(m_1,m_2)}$.  Creation and annihilation
operators need only have one index corresponding to the number of
derivatives acting on a field~$\phi$ \beq [ a_j, a_{j'}^\dag ] =
\delta_{jj'} .  \eeq It is straightforward to show that $V$ from
eq.~(\ref{eq24}) takes the following simple structure on this space
\beq V=\frac{1}{2}\:  \sum_{J=0}^\infty \frac{1}{J+1} \sum_{j,k=0}^J
\;a_j^\dag\,a_{J-j}^\dag\,a_k\,a_{J-k} .  \label{eq42} \eeq
Eigenvalues of $V$ and anomalous dimensions are again connected by
eq.~(\ref{eq25}) and (\ref{eq26}).  It will be quite remarkable to see
what a complicated spectrum is generated by such a seemingly innocent
operator.

The symmetry group of $V$ left over from the full SO(5,1) is SO(2,1)
generated by \beqn D&=&\sum_{j=0}^\infty (j+1)\:  a_{j+1}^\dag\,a_j
\nonumber \\ D^\dag&=&\sum_{j=0}^\infty (j+1)\:  a_{j}^\dag\,a_{j+1}
\label{eq43} \\ S&=&\sum_{j=0}^\infty (j+1/2)\:  a_{j}^\dag\,a_j
\nonumber \eeqn with commutators \beq [S,D]=D , \qquad
[S,D^\dag]=-D^\dag , \qquad [D^\dag,D]=2S .  \label{eq44} \eeq In
order to construct the CIOs, one now introduces the one to one mapping
between a composite operator \beq |\psi\ket=\sum_{\{j_i\}}
c_{j_1\ldots j_n} \:  a_{j_1}^\dag \ldots a_{j_n}^\dag\:|\Omega\ket
\label{eq61} \eeq and a symmetric homogeneous polynomial of degree~$l$
in $n$~variables \beq p_\psi(x_1,\ldots ,x_n) = \sum_{\{j_i\}}
c_{j_1\ldots j_n} \:  x_1^{j_1}\cdot\ldots\cdot x_n^{j_n} .
\label{eq62} \eeq The coefficients $c_{j_1\ldots j_n} $ can be assumed
totally symmetric.  It is easy to see that conformal invariance of
$|\psi\ket$ translates into translation invariance of the polynomial
$p_\psi(x_1,\ldots ,x_n)$ \beq D^\dag |\psi\ket = 0
\qquad\Longleftrightarrow\qquad \left(\frac{\partial}{\partial
x_1}+\ldots+\frac{\partial}{\partial x_n} \right)\:
p_\psi(x_1,\ldots,x_n) = 0 .  \label{eq63} \eeq A basis
$b^{(k)}(x_1,\ldots,x_n)$ for the vector space of symmetric
translation invariant polynomials of degree~$l$ is generated by
products \beq
b^{(k)}(x_1,\ldots,x_n)=\left[q_2(x_1,\ldots,x_n)\right]^{i_2^{(k)}}\cdot
\left[q_3(x_1,\ldots,x_n)\right]^{i_3^{(k)}}\cdot\,\ldots\,\cdot
\left[q_{n-1}(x_1,\ldots,x_n)\right]^{i_{n-1}^{(k)}} , \eeq where \beq
2i_2^{(k)}+3i_3^{(k)}+\ldots+(n-1)i_{n-1}^{(k)}=l ,\qquad
i_j^{(k)}\geq 0 \label{eq7a} \eeq and \beqn
q_m(x_1,\ldots,x_n)&=&\sum_{\pi\in~{\rm S}_n}
s_m(x_{\pi(1)},\ldots,x_{\pi(m)}) , \nonumber\\
s_m(y_1,\ldots,y_m)&=&\prod_{j=1}^m \left(\Big(\sum_{h=1\atop h\neq
j}^m y_h\Big) -(m-1)\,y_j \right) .  \eeqn The different partitions of
$l$ in (\ref{eq7a}) yield all linear independent basis vectors
$b^{(k)}(x_1,\ldots,x_n)$, i.e.  generate ${\cal C}[n,(l;l/2,l/2)]$.
The combinatorial problem of finding all partitions (\ref{eq7a}) is
well--known \cite{bateman53c} and the answer given in terms of a
generating function \beq \sum_{l=0}^\infty \left({\rm dim}\:{\cal
C}[n,(l;l/2,l/2)]\right)\, \cdot x^l = \frac{1}{\prod\limits_{i=2}^n
(1-x^i)} \ \ .  \label{eq45} \eeq For large~$l$ one can prove the
following asymptotic behaviour by expanding the polynomial on the
right hand side of eq.~(\ref{eq45}) and approximating its coefficients
\beq {\rm dim}\:{\cal C}[n,(l;l/2,l/2)] \simeq
\frac{1}{n!\,(n-2)!}\,\left(l+\frac{(n-1)(n+2)}{4}\right)^{n-2} \ \ .
\label{eqasympt} \eeq Thus for $n\geq 4$ the dimensions dim$\:{\cal
C}[n,(l;l/2,l/2)]$ increase quickly with~$l$, making the operator
mixing problem very hard.  The field content of the $N$--vector model
in $4-\epsilon$~dimensions therefore becomes extremely large for many
gradients.

We have used a computer program to generate the polynomials
$b^{(k)}(x_1,\ldots,x_n)$ and thus the corresponding CIOs.  Then a
diagonalization of~$V$ from eq.~(\ref{eq42}) was done in these spaces.
The results for $n\leq 12$, $l\leq 12$ are summed up in table~7.  This
table is not meant to scare the reader, but shall mainly give some
idea of the complexity of the spectrum of anomalous dimensions encoded
by~$V$.  This is of course even more true if one allows for arbitrary
SO(4) spin structure.

Looking at table~7 more closely one finds in particular:
\begin{itemize} \item The lower bounds established in sect.~5,
eq.~(\ref{lowbound}) \beq V\Big|_{{\cal C}[n,(l;l/2,l/2]} \geq
\frac{1}{2} n\,(n-1)-\frac{1}{6} l\,(l+n-1) \eeq and this bound is
strict for $n\geq l$, $l\neq 1$.  \item For $n$ fixed,
$l\rightarrow\infty$ the smallest eigenvalues seem to converge to~0.
This will be explained in the next section and is connected with the
intuitive idea from Parisi that a high spin ``effectively" separates
space--time points \cite{callan73a}.  \item The eigenvalues for $n=4$
and an odd number of gradients~$l$ are just the rational numbers
appearing in the spectrum for $n=3$~fields, plus the additional
eigenvalue~1.  This eigenvalue~1 also appears degenerate for even
larger values of~$l$ (e.g.  twofold degenerate for $l=13$).  We know
no explanation for this feature.  \end{itemize}

\section{Eigenvalues 0 in the spectrum} The case of vanishing
anomalous dimensions in one--loop order is especially interesting
since besides their canonical dimension, the corresponding
eigenoperators are the ``least" irrelevant.  This might be of
importance for operator product expansions on the light cone, see
below.  Besides it is also possible to give a concise classification
of such eigenoperators with eigenvalue~0.

In our previous work \cite{kehrein93a} we already established the
following expression for~$V$ \beqn V&=&\frac{1}{2}\:
\sum_{(L;M_1,M_2)}\sum\limits_{k=0}^{\left[\frac{L}{2}\right]}
(-1)^{L+M_1+M_2}\:\frac{(L-2k)!\,(L-2k+1)!}{(L+1-k)!^2\,k!^2}
\cdot\nonumber \\ &&\qquad\qquad \cdot\:H_{L-2k}^{(M_1,M_2)} ({\rm
ad}\,\vec P)\:\big((\frac{{\rm ad}\,\vec P}{2})^2\big)^k
\:[a^{(0;0,0)\dag}a^{(0;0,0)\dag}]\cdot \nonumber \\
&&\qquad\qquad\cdot\:  H_{L-2k}^{(-M_1,-M_2)} ({\rm ad}\,\vec
K)\:\big((\frac{{\rm ad}\,\vec K}{2})^2\big)^k
\:[a^{(0;0,0)}a^{(0;0,0)}] .  \label{eq100} \eeqn Here ${\rm
ad}\,P_\mu[X]$ denotes the commutator $[P_\mu,X]$ and \beqn \vec
P&=&\big(\partial_{++}-\partial_{--}\,,\,\partial_{+-}+\partial_{-+}\,,\,
i(\partial_{+-}-\partial_{-+})\,,\,-i(\partial_{++}+\partial_{--})\big)
\nonumber \\ K_\mu &=& -P_\mu^\dag .  \eeqn In order to be an
eigenoperator with eigenvalue~0, a conformal invariant operator
$|\psi\ket$ must obviously possess a vanishing matrix element
$<\psi|\,V\,|\psi\ket=0$.  According to (\ref{eq100}) this is
equivalent to $a^{(0;0,0)}\,a^{(0;0,0)}\,|\psi\ket=0$, that is the
composite operator must not contain any factors~$\phi^2$ without
derivatives acting on it \beq V|\psi\ket=0
\quad\Longleftrightarrow\quad a^{(0;0,0)}\,a^{(0;0,0)}\,|\psi\ket=0
\label{eq101} \eeq for CIOs ($\partial_{st}^\dag\psi=0$).

In the case of spatially symmetric and traceless tensors discussed in
the previous section, the above statement can be made more precise.
One easily shows that the action of~$V$ on~$|\psi\ket$ can be
represented on $p_\psi(x_1,\ldots,x_n)$ from eq.~(\ref{eq62}) by \beq
V\:p_\psi(x_1,\ldots ,x_n) = \sum_{{i,j=1}\atop{i<j}}^n
V^{(ij)}\:p_\psi(x_1,\ldots,x_n) , \eeq where \beqn
\lefteqn{V^{(ij)}\:p_\psi(x_1,\ldots,x_n) } \nonumber\\ &=&\int_0^1
du\;\:  p_\psi(x_1,\ldots,x_{i-1},
u\,x_i+(1-u)\,x_j,x_{i+1},\ldots,x_{j-1}, \label{eq102} \\
&&\qquad\qquad u\,x_i+(1-u)\,x_j,x_{j+1},\ldots,x_n) .  \nonumber
\eeqn A sufficient condition for $V\,|\psi\ket=0$ is then \beq \forall
i,j\quad x_i=x_j \quad\Longrightarrow\quad p_\psi(x_1,\ldots,x_n) = 0 .
\label{eq103} \eeq In the appendix we show that this is in fact
also a necessary condition:  The non--local interaction (\ref{eq102})
is equivalent to a $\delta$--interaction potential for a quantum
mechanical system of $n$~bosons living on the twofold covering of the
homogeneous space SO(2,1)/SO(2).  A necessary and sufficient condition
for eigenvalues~0 is that the wave function vanishes if any two
coordinates coincide, which just turns out to be equivalent to
eq.~(\ref{eq103}).

Eq.~(\ref{eq103}) has to be solved for translation invariant
polynomials.  But then the solutions are simply Laughlin's polynomials
\cite{laughlin83a} for bosons \beq p(x_1,\ldots,x_n) =
\left(\prod_{{i,j=1}\atop{i<j}}^n (x_i-x_j)^2 \right)
\:\cdot\:q(x_1,\ldots,x_n) , \label{eq104} \eeq where
$q(x_1,\ldots,x_n)$ can be an arbitrary homogeneous symmetric and
translation invariant polynomial.  Thus one finds CIOs with vanishing
anomalous dimensions for \beqn l&=&n\,(n-1)\ \ \ \ \ \, \qquad
\Big[\;q(x_1,\ldots,x_n)=1\;\Big] \nonumber \\ l&=&n\,(n-1)+2 \qquad
\Big[\;q(x_1,\ldots,x_n)=\sum_{{i,j=1}\atop{i<j}}^n (x_i-x_j)^2\;\Big]
\nonumber \\ l&=&n\,(n-1)+3 \qquad \Big[\;q(x_1,\ldots,x_n) \nonumber
\\ &&\qquad\qquad\qquad\qquad=\sum_{{i,j,k=1}\atop{i<j<k}}^n
(x_i+x_j-2x_k)(x_j+x_k-2x_i)(x_k+x_i-2x_j)\;\Big] \nonumber \\
&\vdots& \nonumber \eeqn The number of eigenvalues zero in ${\cal
C}[n,(l;l/2,l/2)]$ is just the number of linear independent
polynomials $q(x_1,\ldots,x_n)$ in (\ref{eq104}), and this number is
simply ${\rm dim~}{\cal C}[n,(l_0;l_0/2,l_0/2)]$ given by
eq.~(\ref{eq45}) with $l_0=l-n(n-1)$.  The degeneracy in the subspace
of CIOs with vanishing anomalous dimension in one--loop order
therefore becomes arbitrarily large for $l\rightarrow\infty$.  In fact
one has according to eq.~(\ref{eqasympt}) \beq \frac{\Big({\rm
\#~Eigenvalues~0~in~}{\cal C}[n,(l;l/2,l/2)]\Big)}{{\rm dim~} {\cal
C}[n,(l;l/2,l/2)]}\;\;
\stackrel{l\rightarrow\infty}{\longrightarrow}\; 1 \ \ .  \eeq For
large enough~$l$ ``almost all" eigenvalues are zero for any fixed
number of elementary fields~$n$.

Now already Parisi has conjectured that for composite operators like
$\phi\,\partial_{\mu_1}\ldots\partial_{\mu_l}\phi$ in the limit of
large spin $l\rightarrow\infty$, no further subtractions besides the
renormalization of the field~$\phi$ are necessary (compare
ref.~\cite{callan73a}).  An intuitive argument would be to say that a
large angular momentum ``effectively" separates the two space--time
points.  From our results in this section we can extend this statement
to an arbitrary number of space--time points, at least in one--loop
order.  Thereby even higher twist contributions in an operator product
expansion on the light cone are (in one--loop order) ``dominated" by
their canonical scaling behaviour.

\section{Conclusions} In this paper we have investigated the structure
of the one--loop spectrum of anomalous dimensions in the $N$--vector
model in $4-\epsilon$~dimensions.  For a small number of gradients
($l\leq 5$) or elementary fields ($n\leq 3$) in the composite
operators one can derive explicitly the anomalous dimensions of the
scaling eigenoperators.  Neglecting some obvious complications due to
the O($N$) degrees of freedom of the model by either setting $N=1$ or
by considering completely symmetric traceless O($N$) tensors only,
these anomalous dimensions are rational numbers.  Let us mention that
wherever we have compared these results with results obtained by Lang
and R\"uhl in a $1/N$--expansion using operator product expansion
techniques \cite{ruehl92a,ruehl93a,ruehl94a}, we have found agreement.

However the simple structure of the spectrum of anomalous dimensions
does not extend to more complex composite operators ($l>5$ or $n>3$).
The essential reason for this turns out to be the increasing
dimensionality of the spaces of mixing conformal invariant operators
(CIOs), compare e.g.  table~5 or eq.~(\ref{eq45}).  One does not find
a closed expression for the eigenvalues which are in general
non--rational numbers.  This complicated structure even in one--loop
order is quite different from the familiar algebraic structures in two
dimensional conformal field theories.  Clearly life in
$d=4-\epsilon$~dimensions is considerably more complicated than in
$d=2$~dimensions since conformal symmetry yields less stringent
conditions.  In the language of two dimensional conformal field theory
one would say that the field content here in $d=4-\epsilon$~dimensions
is infinite with seemingly no underlying algebraic structure.

In our opinion this is not remedied by the fact that some features in
this complicated spectrum can be understood analytically.  For quite
general classes of CIOs we have been able to work out the smallest
anomalous dimensions in a given space of mixing CIOs explicitly and to
characterize the corresponding scaling eigenoperator (``ground
states").  In particular for a large enough number of gradients for a
fixed number of fields, one finds a highly degenerate subspace of CIOs
with vanishing anomalous dimensions in one--loop order.  This agrees
with Parisi's conjecture that a high spin ``effectively" separates
space--time points in the sense that no additional subtractions are
necessary to renormalize the composite operator.

Finally let us mention some of the remaining problems.  First of all
we would not expect that much more can be learned explicitly about the
eigenvalues in the spectrum of anomalous dimensions beyond the
``ground state" properties that we have discussed in this paper.  In
principle one can of course argue that all the relevant information is
encoded in the relatively simple spectrum--generating operator~$V_N$
anyway.  A remarkable observation in the spectrum is certainly the
infinite number of degenerate eigenvalues, especially with vanishing
anomalous dimension in order~$\epsilon$.  One can always wonder
whether there are deeper physical reasons (symmetries) responsible for
such degeneracies.  We do not consider this very likely and intend to
investigate whether these degeneracies are lifted in two--loop order
in a subsequent publication.

{~}

\noindent The authors would like to thank Y.M.  Pismak and K.~Lang for
valuable discussions.  \newpage \begin{appendix}
\section{Representation as a local interaction problem} A more
intuitive understanding of the spectrum of anomalous dimensions for
spatially symmetric and traceless tensors can be gained by looking
at~$V$ in eq.~(\ref{eq42}) from a different viewpoint:  $V$ can be
regarded as a local interaction potential for a quantum mechanical
system of bosons.  This interpretation is particularly useful for
saying something about the $l\rightarrow\infty$ limit of the spectrum.

As already mentioned in sect.~7, eq.~(\ref{eq102}) unfortunately
corresponds to a non--local interaction.  In order to have a
``physical" local interaction potential we have to work with wave
functions in a suitable two dimensional configuration space~$S$.
Considering the SO(2,1) symmetry of~$V$ it is not surprising that this
space~$S$ is provided by the homogeneous space SO(2,1)/SO(2) --- or to
be precise its twofold covering.  $S$ can be parametrized as the
twofold covering of the upper nappe of a hyperboloid embedded in
$\Re^3$ with metric \beq \left(\begin{array}{ccc} -1&0&0 \\ 0&-1&0 \\
0&0&1 \end{array} \right) .  \label{eq67} \eeq and parametrized by
\beqn y_1&=&\sinh\tau\:\cos\alpha \nonumber \\
y_2&=&\sinh\tau\:\sin\alpha \label{eq69} \\ y_3&=&\cosh\tau .
\nonumber \eeqn The SO(2,1) symmetry group is realized on~$S$ by
generators of rotations (compare ref.~\cite{vilenkin68a}) \beqn
H_+&=&-A_1-i\,A_2= e^{-i\,\alpha}\left(
-\frac{1}{2\sinh\tau}-\frac{\partial}{\partial\tau}+i\,\coth\tau
\frac{\partial}{\partial\alpha}
\right) \nonumber \\ H_-&=&-A_1+i\,A_2= e^{i\,\alpha}\left(
\frac{1}{2\sinh\tau}-\frac{\partial}{\partial\tau}-i\,\coth\tau
\frac{\partial}{\partial\alpha}
\right) \label{eq6a} \\
H_3&=&i\,A_3=i\,\frac{\partial}{\partial\alpha} .  \nonumber \eeqn
corresponding to $D, -D^\dag, S$ from eq.~(\ref{eq44}).

In order to represent the spectrum problem of~$V$ as a quantum
mechanical problem, one maps a composite
operator~$|\psi\ket\;\in\;{\cal C}[n,(l;l/2,l/2)]$ on an $n$--particle
wave function for bosons on~$S$ defined via \beq
|\psi\ket=\sum_{\{j_i\}} c_{j_1\ldots j_n} \:  a_{j_1}^\dag \ldots
a_{j_n}^\dag\:|\Omega\ket \quad\longrightarrow\quad
\Psi(y_1,\ldots,y_n)=\sum_{\{j_i\}} c_{j_1\ldots j_n} \:\:
\sigma_{j_1}(y_1)\,\cdot\ldots\cdot\,\sigma_{j_n}(y_n) .  \label{App1}
\eeq $\sigma_j(y)$ are the one--particle wave functions \beq
\sigma_j(y(\tau,\alpha))=e^{-i\,(j+1/2)\,\alpha}\cdot\,\frac{1}{
\cosh\frac{\tau}{2}}\;
\left(\tanh\frac{\tau}{2}\right)^j \eeq and $0\leq\tau < \infty$,
$0\leq\alpha< 4\pi$ in the notation of eq.~(\ref{eq69}).
$\Psi(y_1,\ldots,y_n)$ has the same transformation properties under
$H_+, H_-, H_3$ as $|\psi\ket$ under $D, -D^\dag, S$.

Notice that the wave functions constructed in this manner do not span
a complete basis on~$S$.  In fact they span a basis in the subspace
with eigenvalue~$1/4$ of the Laplace operator \beq
\Delta_S=-A_1^2-A_2^2+A_3^2 \eeq with $A_1, A_2, A_3$ from
(\ref{eq6a}).  In general the eigenvalues of~$\Delta_S$ are
$(1/4+\rho^2)$ for the odd representations that we are using
($\epsilon=1/2$ in the notation of ref.~\cite{vilenkin68a}).
Therefore the wave functions $\Psi(y_1,\ldots,y_n)$ here have lowest
``kinetic energy" on~$S$.

It turns out that it is just a $\delta$--potential interaction on~$S$
that mimicks the respective action of~$V$ on $|\psi\ket$:  \beq
|\,\psi\ket\stackrel{\rm{eq.~}(\ref{App1})}{\longrightarrow}
\Psi(y_1,\ldots,y_n) \qquad \Longrightarrow \qquad V\,|\,\psi\ket
\stackrel{\rm{eq.~}(\ref{App1})}{\longrightarrow}
H\:\Psi(y_1,\ldots,y_n) \eeq The quantum mechanical Hamiltonian~$H$ is
a sum of two--particle $\delta$--interaction potentials \beqn
h(y_i,y_j)&=&\frac{1}{8\pi}\:\delta(||
y_i(\tau_i,\psi_i)-y_j(\tau_j,\psi_j) ||) \nonumber \\
&=&\frac{1}{8\pi}\:\frac{1}{\sinh\,\tau_i}\:\delta(\tau_i-\tau_j)\:
\delta(\alpha_i-\alpha_j)
\eeqn in the sense that \beqn H\:\Psi(y_1,\ldots,y_n)&\stackrel{\rm
def}{=}& \sum_{i,j=1 \atop i<j}^n \sum_{l_1,l_2=0}^\infty
\sigma_{l_1}(y_i)\,\sigma_{l_2}(y_j)\:  \int_S\:dS(y_i')\,dS(y_j') \\
&&\qquad\times
\overline{\sigma_{l_1}(y_i')}\:\overline{\sigma_{l_2}(y_j')}\,
h(y_i',y_j')\,\Psi(y_1,\ldots,y_i',\ldots,y_j',\ldots,y_n) \nonumber
\eeqn with the induced surface element \beq
\int_S\,dS\big(y(\tau,\alpha)\big)=\int_0^\infty d\tau\:\sinh\,\tau\:
\int_0^{4\pi} d\alpha .  \eeq From this viewpoint it is obvious that a
necessary and sufficient condition for a vanishing eigenenergy of~$H$
is that the wave function $\Psi(y_1,\ldots,y_n)$ vanishes if any two
coordinates coincide.  By virtue of \beq
\Psi(\tau_1,\alpha_1;\ldots;\tau_n,\alpha_n)
=\frac{e^{-\frac{i}{2}(\alpha_1+\ldots+\alpha_n)}}{\cosh\frac{\tau_1}{2}
\cdot\ldots\cdot
\cosh\frac{\tau_n}{2}} \:\cdot\:
p_\psi\left(e^{-i\,\alpha_1}\,\tanh\frac{\tau_1}{2},\ldots,e^{
-i\,\alpha_n}\,
\tanh\frac{\tau_n}{2}\right) \eeq this is equivalent to condition
(\ref{eq103}) in sect.~8.

Finally one can wonder whether there are more common features between
our problem here and the fractional quantum Hall effect (FQHE) besides
the use of Laughlin wave functions.  Loosely speaking this is true to
some extent:  In a certain sense the restriction of wave functions
on~$S$ to those with eigenvalue~$1/4$ of the Laplace
operator~$\Delta_S$ resembles the restriction to the lowest Landau
level in the FQHE.  And instead of fermions in the plane or on a two
dimensional sphere interacting with Coulomb potentials we are
interested in bosons living on the twofold covering of a hyperboloid
with a $\delta$--interaction.  As a consequence of this
$\delta$--interaction the ground state energy vanishes exactly below
the threshold $n\,(n-1)\leq l$ (with the exception of $n\,(n-1)=l+1$).
Similarly the ground state energy per particle vanishes in the FQHE
below a certain filling factor~\cite{hoffmann86a}.  \newpage
{\Huge\underline{\bf{Tables:}}} \begin{table}[h] \begin{center}
\begin{tabular}{|l|l|l|}\hline \multicolumn{3}{|c|}{\bf CIOs with
$n=2$ fields} \\ \hline \# Gradients & O($N$) transformation &
Anomalous dimension $\lambda$ \\ & properties & \\ \hline $l=0$
&Scalar & $\epsilon\,\cdot\,\frac{\displaystyle N+2}{\displaystyle
N+8}+{\rm O}(\epsilon^2)$ \\ & \usebox{\k}\usebox{\k}~~for $N \geq 2$
& $\epsilon\,\cdot\,\frac{\displaystyle 2}{\displaystyle N+8}+{\rm
O}(\epsilon^2)$ \\ $l \geq 1$, odd &
\usebox{\k}\hspace*{-5mm}\raisebox{-4.9mm}{\usebox{\k}}~~for $N \geq
2$ & O($\epsilon^2$) \\ $l\geq 2$, even & Scalar (Notation:
$T^{(l)}$) & O($\epsilon^2$) \\ & \usebox{\k}\usebox{\k}~~for $N \geq
2$ & O($\epsilon^2$) \\ \hline \end{tabular} \end{center} \caption{The
(trivial) spectrum of anomalous dimensions for conformal invariant
operators with $n=2$~fields.  Notice that $T^{(2)}$ is proportional to
the stress tensor.}  \end{table} \begin{table} \begin{center}
\begin{tabular}{|l|l|} \hline \multicolumn{2}{|c|}{\bf CIOs with $n=3$
fields for $N=1$} \\ \hline \# Gradients & Anomalous dimension
$\lambda$ \\ \hline $l=0$ & $\epsilon\,+\, $O($\epsilon^2$) \\ $l\geq
2$ & $\frac{\displaystyle \epsilon}{\displaystyle 3}\,\cdot\, \left(
1+ (-1)^l\,\frac{\displaystyle 2}{\displaystyle l+1}\right)\,
+\,$O($\epsilon^2$) \\ \hline \end{tabular} \end{center} \caption{The
non--zero anomalous dimensions in the spectrum for $n=3$~fields and
$N=1$ component.}  \end{table} \begin{table} \begin{center}
\begin{tabular}{|l|l|l|} \hline \multicolumn{3}{|c|}{\bf CIOs with
$n=3$ fields for $N \geq 2$ components} \\ \hline \# Gradients & O(N)
transformation & Anomalous dimension $\lambda$ \\ & properties & \\
\hline $l=0$ & \usebox{\k}\usebox{\k}\usebox{\k} & $\epsilon\,\cdot\,
\frac{\displaystyle 6}{\displaystyle N+8}\,+\,$O($\epsilon^2$) \\ &
\usebox{\k}~~(two indices & $\epsilon\,+\,$O($\epsilon^2$) \\ &
\hspace*{0.5cm} are contracted) & \\ $l=1$ &
\usebox{\k}\hspace*{-5mm}\raisebox{-4.9mm}{\usebox{\k}}\usebox{\k}&
$\epsilon\,\cdot\,\frac{\displaystyle 3}{\displaystyle
N+8}\,+\,$O($\epsilon^2$) \\ & \usebox{\k} &
$\epsilon\,\cdot\,\frac{\displaystyle N+2}{\displaystyle N+8}\,+\,$
O($\epsilon^2$) \\ $l \geq 2$ & \usebox{\k}\usebox{\k}\usebox{\k} &
$\epsilon\,\cdot\,\frac{\displaystyle 2}{\displaystyle
N+8}\left(1+(-1)^l \frac{\displaystyle 2}{\displaystyle
l+1}\right)+$O($\epsilon^2$) \\ &
\usebox{\k}\hspace*{-5mm}\raisebox{-4.9mm}{\usebox{\k}}\usebox{\k} &
$\epsilon\,\cdot\,\frac{\displaystyle 1}{\displaystyle
N+8}\left(2-(-1)^l \frac{\displaystyle 2}{\displaystyle l+1}\right)
+$O($\epsilon^2$) \\ & \usebox{\k} & Eigenvalues of \\ & &
$\frac{\displaystyle\epsilon}{\displaystyle N+8}\left(
\begin{array}{ll} N+2 & 1+(-1)^l\frac{2}{l+1} \\ (N+2)\cdot (-1)^l
\frac{2}{l+1} & 2+(-1)^l\frac{4}{l+1} \end{array} \right)$ \\ & &
+O($\epsilon^2$) \\ \hline \end{tabular} \caption{The non--zero
anomalous dimensions in the spectrum for $n=3$~fields and $N\geq 2$
components.}  \end{center} \end{table} \renewcommand{\arraystretch}{1}
\begin{table} \begin{center} \begin{tabular}{|l|l|c|l|} \hline
\multicolumn{4}{|c|}{\bf CIOs for $N=1$ or with O($N$) structure
(\ref{largeN3})} \\ \hline \# Gradients & \# Fields & SO(4) spin
structure & Eigenvalues $\alpha$ of $V$ \\ \hline $l=0$ & $n\geq 1$&
(0,0) & $1/2\,n(n-1)$ \\ $l=2$ & $n\geq 2$& (1,1) &
$1/2\,n(n-1)-1/3\,n-1/3$ \\ $l=3$ & $n\geq
3$&(3/2,3/2)&$1/2\,n(n-1)-1/2\,n-1$ \\ $l=4$ & $n\geq 2$&(2,2)
&$1/2\,n(n-1)-3/5\,n+1/5$ \\ & $n\geq 4$&(2,2) &$1/2\,n(n-1)-2/3\,n-2
$ \\ & $n\geq 3$&(2,0),(0,2)&$1/2\,n(n-1)-2/3\,n-1$ \\ & $n\geq
4$&(1,1) & $1/2\,n(n-1)-2/3\,n-2/3$ \\ & $n\geq 4$&(0,0) &
$1/2\,n(n-1)-2/3\,n$ \\ $l=5$ & $n\geq
3$&(5/2,5/2)&$1/2\,n(n-1)-2/3\,n-1/3$ \\ & $n\geq
5$&(5/2,5/2)&$1/2\,n(n-1)-5/6\,n-10/3$ \\ & $n\geq
3$&(5/2,3/2),(3/2,5/2)&$1/2\,n(n-1)-5/6\,n-1/2$ \\ & $n\geq
4$&(5/2,1/2),(1/2,5/2)&$1/2\,n(n-1)-5/6\,n-2$ \\ & $n\geq
5$&(3/2,3/2)&$1/2\,n(n-1)-5/6\,n-5/3$ \\ & $n\geq
4$&(3/2,1/2),(1/2,3/2)&$1/2\,n(n-1)-5/6\,n-7/6$ \\ & $n\geq
5$&(1/2,1/2)&$1/2\,n(n-1)-5/6\,n-2/3$ \\ \hline \end{tabular}
\end{center} \caption{Eigenvalues $\alpha$ of $V$ from
eq.~(\protect\ref{eq24}).  The above list of conformal invariant
operators is complete for $l\leq 5$ gradients.}  \end{table}
\begin{table} \begin{center}
\begin{tabular}{|r|l|c|c|c|r|l|c|c|}\cline{1-4}\cline{6-9} $l$ &
($j_1,j_2,\pi$) & \# CIOs& \# Ground &\hspace*{0.7cm}&
$l$&($j_1,j_2,\pi$)& \# CIOs& \# Ground \\ & & & states & & & & &
states \\ \cline{1-4} \cline{6-9} 0&(0,0,+)&1&1&& 7&(7/2,7/2,+)&4&1 \\
2&(1,1,+)&1&1&& &(7/2,5/2)&3&0 \\ 3&(3/2,3/2,+)&1&1&& &(7/2,3/2)&4&1
\\ 4&(2,2,+)&2&1&& &(7/2,1/2)&2&0 \\ &(2,0) &1&1&& &(5/2,5/2,+)&4&1 \\
&(1,1,+)&1&1&& &(5/2,3/2)&5&1 \\ &(0,0,+)&1&1&& &(5/2,1/2)&3&1 \\
5&(5/2,5/2,+)&2&1&& &(3/2,3/2,+)&6&3 \\ &(5/2,3/2)&1&0&&
&(3/2,3/2,--)&1&0 \\ &(5/2,1/2)&1&1&& &(3/2,1/2)&4&2 \\
&(3/2,3/2,+)&1&1&& &(1/2,1/2,+)&3&2 \\ &(3/2,1/2)&1&1&& 8&(4,4,+)&7&1
\\ &(1/2,1/2,+)&1&1&& &(4,3) &4&0 \\ 6&(3,3,+)&4&1&& &(4,2)&8&1 \\
&(3,2) &1&0&& &(4,1)&3&0 \\ &(3,1)&3&1&& &(4,0)&4&0 \\ &(2,2,+)&3&1&&
&(3,3,+)&8&1 \\ &(2,1)&2&1&& &(3,2)&9&1 \\ &(1,1,+)&4&3&& &(3,1)&8&1
\\ &(0,0,+)&2&2&& &(3,0)&2&0 \\ &&&&&& (2,2,+)&14&3 \\ &&&&&&
(2,2,--)&2&0 \\ &&&&&& (2,1)&9&2 \\ &&&&&& (2,0)&7&2 \\ &&&&&&
(1,1,+)&10&4 \\ &&&&&& (1,1,--)&1&0 \\ &&&&&& (1,0) &2&0 \\ &&&&&&
(0,0,+) &5&3 \\ \cline{1-4}\cline{6-9} \end{tabular} \end{center}
\caption{A complete list of CIOs for $l\leq 8$,
$n\geq[l-1/2\,|j_1-j_2|]$.  CIOs with $j_1\neq j_2$ are only listed
once for $j_1>j_2$.  \#~Ground states is the number of ground state
eigenvalues eq.~(\protect\ref{eq51}) in the spectrum for the
respective quantum numbers.}  \end{table} \begin{table} \begin{center}
\begin{tabular}{|r|l|l|} \hline $l$&($j_1,j_2,\pi$)&Eigenvalue
$\alpha$ of $V$ in eq.  (\ref{eq24}) \\ \hline
4&(2,2,+)&$1/2\,n(n-1)-3/5\,n+1/5$ \\
5&(5/2,5/2,+)&$1/2\,n(n-1)-2/3\,n-1/3$ \\
&(5/2,3/2)&$1/2\,n(n-1)-5/6\,n-1/2$ \\
6&(3,2)&$1/2\,n(n-1)-14/15\,n-7/5$ \\
&(2,1)&$1/2\,n(n-1)-14/15\,n-2/5$\\
&(1,1,+)&$1/2\,n(n-1)-14/15\,n+8/15$\\
7&(3/2,3/2,--)&$1/2\,n(n-1)-7/6\,n-4/3$\\
&(3/2,1/2)&$1/2\,n(n-1)-11/10\,n-1/2$ \\
&(3/2,1/2)&$1/2\,n(n-1)-7/6\,n-19/30$ \\
&(1/2,1/2,+)&$1/2\,n(n-1)-11/10\,n+1/5$\\
8&(4,0)&$1/2\,n(n-1)-6/5\,n+3/5$\\ &(4,0)&$1/2\,n(n-1)-4/3\,n-2/3 $\\
&(1,1,--)&$1/2\,n(n-1)-4/3\,n-5/3$ \\
&(1,0)&$1/2\,n(n-1)-19/15\,n-14/15$\\
&(1,0)&$1/2\,n(n-1)-4/3\,n-6/5$\\ &(0,0,+)&$1/2\,n(n-1)-6/5\,n+8/5$\\
&(0,0,+)&$1/2\,n(n-1)-19/15\,n-7/15$\\ \hline \end{tabular}
\end{center} \caption{Linear eigenvalues $\alpha$ different from the
ground state eigenvalues for $l\leq 8$, $n\geq[l-1/2\,|j_1-j_2|]$.
All these eigenvalues are non--degenerate.}  \end{table}
\begin{table}[p] \begin{center} {\tiny
\begin{tabular}{|rl||r|r|r|r|r|r|r|r|r|r|r|} \hline
\multicolumn{13}{|c|}{~~} \\ \multicolumn{13}{|c|}{\normalsize\bf CIOs
in ${\cal C}[n,(l;l/2,l/2)]$} \\ \multicolumn{13}{|c|}{~~} \\ \hline
&$n$&2&3&4&5&6&7&8&9&10&11&12 \\ $l$&& & & & & & & & & & & \\ \hline
\hline 0&& 1& 3& 6& 10& 15& 21& 28& 36& 45& 55& 66 \\ \hline 1&& & & &
& & & & & & & \\ \hline 2&& 0& 5/3&13/3& 8& 38/3& 55/3& 25&
98/3&124/3& 51& 185/3 \\ \hline 3&& & 1/2 & 3 & 13/2& 11& 33/2& 23&
61/2& 39& 97/2& 59 \\ \hline 4&&0 & 7/5& 4/3 & 14/3 & 9 & 43/3 & 62/3
& 28 & 109/3& 137/3& 56 \\ && & &19/5&36/5 &58/5 & 17 &117/5
&154/5&196/5&243/5 & 59 \\ \hline 5&& & 2/3& 3& 5/2 & 20/3 & 71/6 & 18
& 151/6& 100/3& 85/2 & 158/3 \\ && & & &19/3 & 32/3& 16 & 67/3 &89/3 &
38 &142/3 &173/3 \\ \hline 6&&0 & 0 & 1.49& 4.51& 4 & 9 & 15 &22 & 30
& 39 & 49 \\ && & 9/7 & 2.01&5.11 &8.49 &13.45 & 19.42&26.38 &34.34&
43.31 &53.28 \\ && & &3.54&6.77 &9.25 &14.40 & 20.56&27.71 &35.86 & 45
&55.14 \\ && & & & &11.01&16.24 & 22.47&29.71 &37.95 & 47.20&57.45 \\
\hline 7 && & 3/4& 1 & 2.76 & 6.69& 35/6 & 35/3 & 37/2& 79/3& 211/6&
45 \\ && & &3 & 3.96 &7.94 & 11.61&17.52&24.42&32.32&41.21&51.10 \\ &&
& & & 6.24 &10.47&12.94&18.95&25.98&34 &43.03&53.06 \\ && & & & &
&15.70&21.93&29.16&37.38&46.61&56.85 \\ \hline 8&&0& 0 & 0.34&2.11&
4.33& 9.10 & 8 & 44/3 & 67/3 & 31 & 122/3 \\ && & 11/9&1.57&3
&5.77&10.42&14.85&21.57&29.27&37.95&47.62 \\ && & &
2.03&4.40&$\ldots$&$\ldots$&$\ldots$&$\ldots$&$\ldots$&$\ldots$&$\ldots$\\
&& & & 3.38&5.09&10.66&15.80&21.95&29.10&37.27&46.44&56.62 \\ & & & &
&6.52& & & & & & & \\ & & & & & & \#6 &\#6 &\#7 &\#7 &\#7 &\#7 &\#7 \\
\hline 9&& & 0 & 1 & 1.66 & 3.64 & 6.23 & 11.84 & 21/2 &18 &53/2 &36
\\ && & 4/5&5/3&2.93&5.28&8.20&13.77&18.44 &26.02 &34.59 &44.15 \\ &&
& & 3 &3.85 &${\ldots}$ &${\ldots}$ &${\ldots}$ &
${\ldots}$&${\ldots}$&${\ldots}$&${\ldots}$ \\ && & & & 4.39 & 10.34 &
15.51 & 21.67 &28.83 &37.00 &46.17 &56.33 \\ && & & & 6.18& & & & & &
& \\ && & & & & \#6 &\#7 & \#7 &\#8 &\#8 &\#8 &\#8 \\ \hline 10&&0& 0
& 0.36 & 0.92 &3.08 &5.49 &8.47 &14.92 &40/3 &65/3 &31 \\ && &13/11&1
&2.16 &4.19 &7.43 &10.89 &17.28 &22.35 &30.78 &40.19 \\ && & &1.60
&$\ldots$&$\ldots$&$\ldots$&$\ldots$&$\ldots$&$\ldots$&$\ldots$&$\ldots$
\\ && & &2.04 &6.37 &10.45 &15.54 &21.65 &28.77 &36.89 &46.03 &56.17
\\ && & &3.29 & & & & & & & & \\ && & & &\#7 &\#9 &\#10 &\#11 &\#11
&\#12 &\#12 &\#12 \\ \hline 11&& &0
&1/2&1.66&2.59&4.90&7.66&11.03&18.32&33/2&77/3 \\ && &5/6&1
&2.08&3.76&6.53&10.16&13.90&21.14&26.60&35.86 \\ && &
&5/3&$\ldots$&$\ldots$&$\ldots$&$\ldots$&$\ldots$&$\ldots$&$\ldots$&$\ldots$
\\ && & &3 &6.14&10.26&15.38&21.50&28.62&36.75&45.87&56.01 \\ && & & &
& & & & & & & \\ && & & & \#7& \#9 &\#11 &\#12 &\#13 & \#13 & \#14 &
\#14 \\ \hline 12&&0&0 &0 &0.93 &1.70 &4.33 &7.04 &10.17
&13.93&22.06&20 \\ && &0 &0.37&1.65 &3.12 &5.51 &8.83 &13.90 &17.25
&25.33&31.17 \\ &&&15/13&$\ldots$&$\ldots$&$\ldots$&$\ldots$&
$\ldots$&$\ldots$&$\ldots$&$\ldots$&$\ldots$ \\ && &
&3.22&6.26&10.32&15.39&21.47&28.57&36.67&45.78&55.89 \\ && & & & & & &
& & & & \\ && & &\#7 &\#10&\#14&\#16&\#18&\#19&\#20&\#20&\#21 \\
\hline \end{tabular} }\end{center} \caption{Spectrum of $V$ from eq.
(\protect\ref{eq42}) for $n,l\leq 12$.  Numbers with decimal points
are approximate numerical results and no rational numbers, otherwise
the given eigenvalues are exact.  \#$\ldots$ denotes the dimension of
the respective space ${\cal C}[n,(l;l/2,l/2)]$ if not all eigenvalues
are written explicitly.}  \end{table} \end{appendix} 


\newpage
\begin{thebibliography}{10}

\bibitem{callan73a} C.G.  Callan and D.J.  Gross, Phys.  Rev.
D8~(1973)~4383

\bibitem{castilla93a} G.~E.  Castilla and S.~Chakravarty, Phys.  Rev.
Lett.~71~(1993)~384

\bibitem{bateman53c} A.~Erd{\'e}lyl, W.~Magnus, F.~Oberhettinger and
F.G.  Tricomi, Higher Transcendental Functions, vol.~3 (McGraw-Hill,
New York, 1953)

\bibitem{hoffmann86a} K.H.  Hoffmann and B.~Neudecker, Z.  Phys.
B62~(1986)~279

\bibitem{kehrein93a} S.K.  Kehrein, F.J.  Wegner and Y.M.  Pismak,
Nucl.  Phys.  B402~(1993)~669

\bibitem{kravtsov88a} V.E.  Kravtsov, I.V.  Lerner and V.I.  Yudson,
Zh.  Eksp.  Teor.  Fiz.~94~(1988)~255; Sov.  Phys.
JETP~67~(1988)~1441

\bibitem{kravtsov89a} V.E.  Kravtsov, I.V.  Lerner and V.I.  Yudson,
Phys.  Lett.  A~134~(1989)~245

\bibitem{ruehl92a} K.~Lang and W.~R{\"u}hl, Nucl.  Phys.
B377~(1992)~371

\bibitem{ruehl93a} K.~Lang and W.~R{\"u}hl, Nucl.  Phys.
B400~(1993)~597

\bibitem{ruehl94a} K.~Lang and W.~R{\"u}hl, Preprint
HEP--TH~9401116~(1994) and private communication.

\bibitem{laughlin83a} R.B.  Laughlin, Phys.  Rev.
Lett.~50~(1983)~1395

\bibitem{mack77a} G.~Mack, Commun.  math.  Phys.~53~(1977)~155

\bibitem{mall92a} H.~Mall and F.J.  Wegner, Nucl.  Phys.
B393~(1993)~495

\bibitem{vilenkin68a} N.J.  Vilenkin, Special Functions and the Theory
of Group Representations (American Mathematical Society, Providence
Rhode Island, 1968)

\bibitem{wegner90a} F.J.  Wegner, Z.  Phys.  B78~(1990)~33

\bibitem{wegner91a} F.J.  Wegner, Nucl.  Phys.  B354~(1991)~441

\bibitem{zimmermann73a} W.~Zimmermann, Ann.  Phys.~77~(1973)~536

\end{thebibliography}
\end{document}